\documentclass[a4paper,
               keeplastbox,   
               ]{jacow}
\usepackage{pdfpages,multirow,ragged2e} %
\makeatletter%
	\ifboolexpr{bool{xetex}}
	 {\renewcommand{\Gin@extensions}{.pdf,%
	                    .png,.jpg,.bmp,.pict,.tif,.psd,.mac,.sga,.tga,.gif,%
	                    .eps,.ps,%
	                    }}{}
\makeatother

\ifboolexpr{bool{xetex} or bool{luatex}} 
 {}                                      
 {\usepackage[utf8]{inputenc}}           

\usepackage[USenglish]{babel}

\ifboolexpr{bool{jacowbiblatex}}%
 {%
  \addbibresource{jacow-test.bib}
  \addbibresource{biblatex-examples.bib}
 }{}
\begin{document}

\title{Potential-well bunch lengthening in electron storage rings}

\author{Demin Zhou\thanks{dmzhou@post.kek.jp}, Gaku Mitsuka, Takuya Ishibashi,\\ 
        KEK Accelerator Laboratory, 1-1 Oho, Tsukuba 305-0801, Japan \\
		Karl Bane, SLAC National Accelerator Laboratory, Menlo Park, California 94025, USA 
  }
	
\maketitle

\begin{abstract}
   The cubic equation derived by B. Zotter has been popularly used for electron storage rings to describe the scaling law of potential-well bunch lengthening. This equation has also often been used to calculate the effective impedance when the bunch lengthening is measured or simulated. This paper discusses the validity of Zotter’s equation and presents an alternative but self-consistent equation for potential-well bunch lengthening. Its applications to predicting bunch lengthening and extracting effective impedance from bunch length measurements are also addressed.
\end{abstract}

\section{Introduction}

Individual electrons oscillate around a fixed point in the longitudinal phase space in electron storage rings. The RF system creates a potential well that confines the beam in a bucket. When the oscillations are of small amplitude, the motion of the electrons is linear. Using the longitudinal coordinate $z$ and momentum deviation $\delta$, the linear equations of motion are\cite{WolskiBook2014}
\begin{equation}
    \frac{dz}{ds}=-\eta \delta, \quad \frac{d\delta}{ds}=-\frac{eV_\text{rf}\omega_\text{rf}\cos \phi_\text{s}}{ECc}z.
    \label{eq:motion1}
\end{equation}
The following quantities are used: the electron charge $e$, the speed of light $c$, the slip factor $\eta$, the RF voltage $V_\text{rf}$, the RF frequency $\omega_\text{rf}$, the electron energy $E$, the circumference of the ring $C$ and the synchronous phase $\phi_\text{s}$. We assume that the reference particle is moving at the speed of light, $c$. In the presence of radiation damping and quantum excitation, the beam reaches a stable equilibrium distribution in the longitudinal phase space, which is Gaussian\cite{ChaoBook2020}. The momentum spread $\sigma_{\delta 0}$ in an electron storage ring at equilibrium is determined by the beam energy and the bending radii of the dipole magnets but is not affected by the RF parameters $V_\text{rf}$ and $\omega_\text{rf}$ (see Eq.~(7.94) of~\cite{WolskiBook2014}). Radiation effects cause the beam to bunch, with an rms length of $\sigma_\text{z0}=-c\eta\sigma_{\delta 0}/\omega_\text{z0}$. From Eqs.~(\ref{eq:motion1}), the synchrotron frequency $\omega_\text{z0}$ is defined by
\begin{equation}
    \omega_\text{z0}\equiv
    \text{sgn}[\eta] \sqrt{-\frac{ecV_\text{rf}\eta\omega_\text{rf}\cos \phi_\text{s}}{EC}},
    \label{eq:freq1}
\end{equation}
where $\text{sgn}[]$ is the sign function. We define $\omega_\text{z0}$ to have the opposite sign of $\eta$ (this is the choice of the SAD code\cite{SADpage}) so that $\sigma_\text{z0}$ is always positive.

The longitudinal wakefields deform the potential well created by the RF system. As a result, the equations of motion are modified to:
\begin{equation}
    \frac{dz}{ds}=-\eta \delta, \quad \frac{d\delta}{ds}=
    \frac{\omega_\text{z0}^2}{\eta c^2}z - F(z,s).
    \label{eq:motion2}
\end{equation}
The wakefield term $F(z,s)$ is calculated from the convolution of charge density and longitudinal wake function
\begin{equation}
    F(z,s)=I_n \int_{-\infty}^\infty W_\parallel(z-z')\lambda(z',s) dz',
    \label{eq:Wake_force1}
\end{equation}
with the scaling factor $I_n=Ne^2/(EC)=I_b/(c(E/e))$, the line density distribution $\lambda(z,s)$, and the wake function $W_\parallel(z)$. Here, $N$ is the bunch population, and $I_b=Ne/(C/c)$ is the bunch current. The wake function and the corresponding impedance are Fourier transforms of each other: $W_\parallel(z)=\frac{c}{2\pi}\int_{-\infty}^\infty Z_\parallel(k) e^{ikz} dk$ and $Z_\parallel(k)=\frac{1}{c}\int_{-\infty}^\infty W_\parallel(z) e^{-ikz} dz$ with $k\equiv \omega/c$. Using the impedance, $F(z,s)$ can be rewritten as
\begin{equation}
    F(z,s)=\frac{cI_n}{2\pi}
    \int_{-\infty}^\infty dk
    Z_\parallel(k) \tilde{\lambda}(k,s) e^{ikz},
    \label{eq:Wake_force2}
\end{equation}
with the beam spectrum $\tilde{\lambda}(k,s)=\int_{-\infty}^\infty \lambda(z,s) e^{-ikz} dz$. Since $W_\parallel(z)$ and $\lambda(z,s)$ are real, the reality condition applies: $\tilde{\lambda}^*(k,s)=\tilde{\lambda}(-k,s)$ and $Z_\parallel^*(k)=Z_\parallel(-k)$.

\section{The cubic equation for bunch lengthening}

In\cite{zotter1981potential}, Zotter solved the equations of motion for a single particle as described by Eqs.~(\ref{eq:motion2}). The leading-order perturbation from the longitudinal wakefields was found to shift the incoherent synchrotron frequency. For electron storage rings, Eq.~(5.10) of \cite{zotter1981potential} is of interest to us:
\begin{equation}
    \omega_\text{z}^2=\omega_\text{z0}^2
    \left( 1 - \xi Z_1 \right),
    \label{eq:freshift1}
\end{equation}
with $\xi=\eta \omega_0^2I_b/(2\pi \omega_\text{z0}^2(E/e))$ and
\begin{equation}
    Z_1(\sigma_\text{z})=-\frac{\sqrt{2\pi}c^3}{\omega_\text{0}^3\sigma_\text{z}^3}
    \text{Im} \left( \frac{Z_\parallel}{n} \right)^{m=1}_\text{eff}.
\end{equation}
Here $\omega_\text{z}$ is the incoherent synchrotron frequency after bunch lengthening, $\sigma_\text{z}$ is the rms length of the lengthened bunch, $\omega_0=2\pi c/C$ is the revolution frequency, and $\left( Z_\parallel/n \right)^{m=1}_\text{eff}$ is the effective impedance for azimuthal mode $m=1$. The effective impedance is a quantity used in collective instability theories\cite{chao1993physics}. It is defined as follows:
\begin{equation}
    \left( \frac{Z_\parallel}{n} \right)_\text{eff}^{m}=
    \frac{\sum\limits_{p=-\infty}^\infty \frac{Z_\parallel(\omega_{mp})}{p}h_m(\omega_{mp})}{\sum\limits_{p=-\infty}^\infty h_m(\omega_{mp})},
    \label{eq:effective_impedance_definition}
\end{equation}
where $n=\omega/\omega_0$, $\omega_{mp}=p\omega_0+m\omega_\text{z}$ the mode frequencies. The power density of the $m$-th mode is defined as
\begin{equation}
    h_m(\omega)=\tilde{\lambda}_m(\omega)\cdot \tilde{\lambda}_m^*(\omega),
\end{equation}
with $\tilde{\lambda}_m(\omega)$ the Fourier transform of the line density of the $m$-th azimuthal mode. It was suggested in \cite{zotter1981potential} that, when using Eq.~(\ref{eq:effective_impedance_definition}) to calculate $Z_1$, $\omega_{mp}$ should be evaluated with $\omega_\text{z}=0$.

Usually, the ring length is much larger than the bunch length, suggesting that the sampling frequency step $\omega_0$ in Eq.~(\ref{eq:effective_impedance_definition}) is much smaller than the typical frequency of the beam spectrum, that is, $c/\sigma_\text{z0}$. With this condition, the summations over $p$ in Eq.~(\ref{eq:effective_impedance_definition}) can be fairly replaced by integrals, yielding
\begin{equation}
    \left( \frac{Z_\parallel}{n} \right)_\text{eff}^{m}=
    \frac{\int_{-\infty}^\infty Z_\parallel(\omega)\frac{\omega_0}{\omega}h_m(\omega)d\omega}{\int_{-\infty}^\infty h_m(\omega)d\omega}.
    \label{eq:effective_impedance_definition_integral}
\end{equation}
For a Gaussian bunch, the power densities $h_m(\omega)$ for $m$-mode is given by
\begin{equation}
    h_m(\omega)=(\omega\sigma_\text{z}/c)^{2m}e^{-\omega^2\sigma_\text{z}^2/c^2}.
    \label{eq:powerdensity1}
\end{equation}
For the case of potential-well bunch lengthening, $\sigma_\text{z}$ can be replaced by $\sigma_\text{z0}$ so that $\left( Z_\parallel/n \right)^{m=1}_\text{eff}$ is a constant machine parameter independent of the bunch current. Note that the effective impedance depends on the beam properties, but the original impedance $Z_\parallel(\omega)$ only depends on the beam's surroundings~\cite{chao1993physics}.

For electron storage rings, the energy spread is constant below the microwave instability threshold, resulting in
\begin{equation}
    \sigma_\text{z}\omega_\text{z}=\sigma_\text{z0}\omega_\text{z0}=-c\eta\sigma_{\delta 0}.
    \label{eq:cond1}
\end{equation}
Combining Eqs.~(\ref{eq:freshift1}) and (\ref{eq:cond1}), we arrive at Zotter's cubic equation for potential-well bunch lengthening:
\begin{equation}
    x^3-x+\frac{cI_b}{\kappa\eta\omega_0\sigma_\text{z0}\sigma_{\delta 0}^2(E/e)}
    \text{Im} \left( \frac{Z_\parallel}{n} \right)^{m=1}_\text{eff}
    =0,
    \label{eq:cubic1}
\end{equation}
where $x=\sigma_\text{z}/\sigma_\text{z0}$ and $\kappa=\sqrt{2\pi}$. Note that from Eq.~(\ref{eq:freshift1}) to Eq.~(\ref{eq:cubic1}), $\xi Z_1(\sigma_\text{z})$ is replaced by $\xi Z_1(\sigma_\text{z0})/x^3$.

There are alternative ways of deriving the cubic equation similar to Eq.~(\ref{eq:cubic1}) (for example, see Refs.\cite{chao1993physics, ng2006physics, chao2022special}). Here, we present a simple way that does not need the effective impedance defined by Eq.~(\ref{eq:effective_impedance_definition}). From Eqs.~(\ref{eq:motion2}), we can obtain the second-order differential equation of longitudinal coordinate as
\begin{equation}
    \frac{d^2z}{ds^2} = -\frac{\omega_\text{z0}^2}{c^2} z
    + \eta F(z,s)
    \label{eq:motion3}
\end{equation}
We assume the incoherent synchrotron frequency with wakefield perturbation is given by
\begin{equation}
    \frac{d^2z}{ds^2} = -\frac{\omega_\text{z}^2}{c^2} z.
    \label{eq:motion4}
\end{equation}
To find the explicit form of $\omega_\text{z}$, we can approximate $F(z,s)$ by taking the leading term that is linear to $z$. To do so, we take the derivative over $z$ on the right side of Eq.~(\ref{eq:Wake_force2}) and then average over the line density:
\begin{equation}
    \begin{split}
        F_1 & =\int_{-\infty}^\infty dz \lambda(z,s) \frac{\partial F(z,s)}{\partial z}\\
            & = \frac{icI_n}{2\pi}
                \int_{-\infty}^\infty dk k Z_\parallel (k) \tilde{\lambda}(k,s) \tilde{\lambda}^*(k,s).
    \end{split}
\end{equation}
From the reality condition of the impedance and beam spectrum, it follows that only the imaginary part of $Z_\parallel(k)$ contributes to $F_1$. Taking a Gaussian bunch and pure inductive impedance $Z_\parallel (k) = -ikcL$, one can obtain
\begin{equation}
    F_1=\frac{cI_n}{2\pi} \frac{\sqrt{\pi}cL}{2\sigma_\text{z}^3}.
    \label{eq:f1}
\end{equation}
Alternatively, for an absolute impedance, the effective inductance can be defined as follows\cite{bane1990bunch}:
\begin{equation}
    L_\text{eff} = \frac{1}{c}
    \frac{\int_{-\infty}^\infty dk k Z_\parallel (k) \tilde{\lambda}(k,s) \tilde{\lambda}^*(k,s)}{\int_{-\infty}^\infty dk k^2 \tilde{\lambda}(k,s) \tilde{\lambda}^*(k,s)}.
\end{equation}
For a Gaussian bunch, it becomes
\begin{equation}
    L_\text{eff} = \frac{2i\sigma_\text{z}^3}{\sqrt{\pi}c}
    \int_{-\infty}^\infty dk k Z_\parallel (k) \tilde{\lambda}(k,s) \tilde{\lambda}^*(k,s),
\end{equation}
which is equivalent to
\begin{equation}
    L_\text{eff} = -\frac{1}{\omega_0} \text{Im} \left( \frac{Z_\parallel}{n} \right)_\text{eff}^{m=1}
\end{equation}
with the effective impedance calculated by Eqs.~(\ref{eq:effective_impedance_definition_integral}) and (\ref{eq:powerdensity1}). The concept of effective inductance was first introduced by K. Bane in\cite{bane1990bunch}.

If we replace $F(z,s)$ with $F_1z$ in Eq.~(\ref{eq:motion3}) and compare the resulting equation with Eq.~(\ref{eq:motion4}), we can obtain
\begin{equation}
    \frac{\omega_\text{z}^2}{c^2} = \frac{\omega_\text{z0}^2}{c^2} - \frac{c^2\eta I_n L_\text{eff}}{4\sqrt{\pi}\sigma_\text{z}^3}.
\end{equation}
Using the relation given in Eq.~(\ref{eq:cond1}), we arrive at a cubic equation of bunch lengthening factor $x$:
\begin{equation}
    x^3-x-\frac{cI_b}{\kappa\eta\sigma_\text{z0}\sigma_{\delta 0}^2(E/e)}
    L_\text{eff}
    =0,
    \label{eq:cubic2}
\end{equation}
with $\kappa=4\sqrt{\pi}$. It can be seen that Eqs.~(\ref{eq:cubic1}) and (\ref{eq:cubic2}) are identical except for the constant $\kappa$ in the denominator. We conclude that the factor of $\sqrt{2\pi}/(\omega_0\tau)^3$ for $Z_1$ in Eq.~(5.9) of\cite{zotter1981potential} is incorrect and should be replaced by $\sqrt{\pi}/(2(\omega_0\tau)^3)$. This conclusion will be justified in the following section.

The cubic equations (\ref{eq:cubic1}) and (\ref{eq:cubic2}) have been widely used in electron storage rings. During the design and construction phases, an impedance database can be created and used to calculate the effective impedance. This can then be used to predict the potential-well bunch lengthening using Eq.~(\ref{eq:cubic2}). During beam commissioning phases, beam-based measurements are typically performed to extrapolate the effective impedance or effective inductance, which is used to verify the accuracy of the impedance database. A comprehensive review can be found in \cite{smaluk2018impedance}. The author of \cite{smaluk2018impedance} suggests that Eq.~(\ref{eq:cubic2}) is a model more consistent with the prediction of the Haissinski equation \cite{haissinski1973exact} and with the measurements in the electron storage rings. However, it is important to emphasize that this consistency is conditional on the following factors:
(1) The total longitudinal impedance of the ring can be well approximated by a pure inductance, $Z_\parallel (k) = -ikcL$, with the impact of its real part negligible.
(2) The potential well, which is distorted by wakefields, remains well quadratic so that the lengthened bunch is close to Gaussian. 
These factors allow us to replace $\eta F$ in Eq.~(\ref{eq:motion3}) with $\eta F_1z$, where $F_1$ is given by Eq.~(\ref{eq:f1}). This leads to the cubic equation (\ref{eq:cubic2}). Typically, important sources of inductive impedance include small 3D discontinuities, such as obstacles or cavities with longitudinal dimensions comparable to or smaller than the rms bunch length ($\sigma_\text{z0}$), on beam pipes; small angle transitions, such as collimators in colliders or tapers sandwiching small-gap insertion devices in light sources, in flat chambers. Small 3D discontinuities from the flanges and bellows are common in all storage rings.

The cubic equation (\ref{eq:cubic2}) may not apply well in some cases. One example is damping rings, where small-angle transitions are usually unnecessary. A good example is the SLC damping ring with the improved chamber\cite{bane1993impedance}, where the resistive part dominates the total impedance\cite{warnock2018numerical}. Another example is future circular colliders (FCCs), where the resistive-wall impedance is dominant because of their large circumferences.

\section{A self-consistent equation for bunch lengthening}

For electron storage rings, the circulating beam can be modeled as a continuous distribution $\psi(z,\delta,s)$, with its evolution governed by the Vlasov-Fokker-Planck (VFP) equation~\cite{frank2020linear}. Specifically, when considering the synchrotron motion, the VFP equation is~\cite{jowett1987introductory}
\begin{equation}
    \frac{\partial\psi}{\partial s}
    +\frac{d z}{d s} \frac{\partial \psi}{\partial z}
    +\frac{d \delta}{d s} \frac{\partial \psi}{\partial \delta}
    =\frac{2}{ct_d} \frac{\partial}{\partial \delta}
    \left[ \delta\psi + \sigma_{\delta0}^2 \frac{\partial \psi}{\partial \delta} \right],
    \label{eq:vfp_equation}
\end{equation}
with $t_d$ the longitudinal damping time. The equations of motion are given by Eqs.~(\ref{eq:motion2}), with the line density distribution $\lambda(z,s)=\int_{-\infty}^\infty \psi(z,\delta,s)d\delta$.

The VFP equation has an $s$-independent stationary solution $\psi_0(z,\delta)$ in the form of $\psi_0(z,\delta)=\hat{\psi}_0(\delta) \lambda_0(z)$ below the microwave instability threshold. The momentum distribution $\hat{\psi}_0(\delta)$ is Gaussian with rms spread $\sigma_{\delta 0}$, and the spatial distribution satisfies
\begin{equation}
    \frac{d\lambda_0(z)}{dz}+
    \left[ \frac{z}{\sigma_\text{z0}^2}-\frac{1}{\eta\sigma_{\delta0}^2} F_0(z) \right]
    \lambda_0(z)
    =0.
    \label{eq:de_Haissinski}
\end{equation}
Here $F_0(z,s)$ is from Eq.~(\ref{eq:Wake_force1}) with $\lambda(z,s)$ replaced by the equilibrium distribution $\lambda_0(z)$. The solution of Eq.~(\ref{eq:de_Haissinski}) is the so-called Haissinski equation~\cite{haissinski1973exact}
\begin{equation}
    \lambda_0(z)=
    A e^{-\frac{z^2}{2\sigma_\text{z0}^2}-\frac{I}{\sigma_\text{z0}}\int_z^{\infty}dz' \mathcal{W}_\parallel(z')},
    \label{eq:Haissinski_equation_explicit}
\end{equation}
with the new scaling parameter $I=I_n\sigma_\text{z0}/(\eta\sigma_{\delta0}^2)$ and the wake potential of the bunch
\begin{equation}
    \mathcal{W}_\parallel(z)=\int_{-\infty}^\infty W_\parallel(z-z')\lambda_0(z') dz',
\end{equation}
With the synchrotron tune defined by $\nu_\text{z0}=\omega_\text{z0}/\omega_0$, there is $I=-Ne^2/(2\pi\nu_\text{z0}E\sigma_{\delta0})$~\cite{oide1990long, warnock2018numerical}. The stability of the Hassinski equation is beyond the scope of this paper, and the reader is referred to~\cite{cai2011linear} and references therein.

Integrating over $z$ on both sides of Eq.~(\ref{eq:de_Haissinski}) and recognizing that the center of mass of the bunch is
\begin{equation}
    z_c=\int_{-\infty}^\infty z\lambda_0(z) dz,
    \label{eq:zc}
\end{equation}
we obtain $z_c=I\sigma_\text{z0}\kappa_\parallel$, with the well-known loss factor
\begin{equation}
    \kappa_\parallel(I)=\int_{-\infty}^\infty dz \lambda_0(z)\mathcal{W}_\parallel(z).
    \label{eq:lossfactor}
\end{equation}
In terms of impedance, it is
\begin{equation}
    \kappa_\parallel=\frac{c}{\pi} \int_0^\infty \text{Re}[Z_\parallel(k)] h(k) dk
\end{equation}
with the spectral power density $h(k)=\tilde{\lambda}_0(k)\tilde{\lambda}_0^*(k)$.

Given the bunch profile $\lambda_0(z)$, by definition, the rms bunch length is calculated by
\begin{equation}
    \sigma_\text{z}^2=\int_{-\infty}^\infty
    (z-z_c)^2\lambda_0(z)dz
    =\int_{-\infty}^\infty
    z^2\lambda_0(z)dz-z_c^2.
\end{equation}
We show how to calculate $\sigma_\text{z}$ from Eq.~(\ref{eq:de_Haissinski}). We can obtain three terms by multiplying $z$ on both sides of this equation and performing integration over $z$. The first term is a constant -1 (consider that $\lambda_0(z)$ decays exponentially as $e^{-z^2/(2\sigma_\text{z0}^2)}$ when $z\rightarrow\pm \infty$, according to the Haissinski equation). The second term equals $\sigma_\text{z}^2+z_c^2$. The third term is an integration that contains the wake function. Combining the three terms, we can arrive at an equation that describes the potential-well bunch lengthening
\begin{equation}
    x^2-1-\frac{cI}{2\pi\sigma_\text{z0}} Z_\parallel^\text{eff}(x)=0,
    \label{eq:PW_Lengthening_from_Haissinski}
\end{equation}
where $Z_\parallel^\text{eff}$ is formulated by
\begin{equation}
    Z_\parallel^\text{eff} =
    \frac{2\pi}{c} \int_{-\infty}^\infty dz (z-z_c) \lambda_0(z)
    \mathcal{W}_\parallel(z).
    \label{eq:Zeff1}
\end{equation}
In terms of impedance, it is equivalent to
\begin{equation}
        Z_\parallel^\text{eff}=
        -\int_{-\infty}^\infty dk Z_\parallel(k) \tilde{\lambda}_0(k)
        \left[ i\frac{d}{dk}\tilde{\lambda}_0^*(k) +
        z_c \tilde{\lambda}_0^*(k) \right].
    \label{eq:Zeff2}
\end{equation}
Here, we define $Z_\parallel^\text{eff}$ as an effective impedance, which is always real but not complex, to indicate bunch lengthening. Equation (\ref{eq:PW_Lengthening_from_Haissinski}) shows that the term $Z_\parallel^\text{eff}$ is a quadratic function of $x$, where $x$ is a function of the normalized current $I$ (i.e., $x=x(I)$). Since Eq.~(\ref{eq:PW_Lengthening_from_Haissinski}) is derived from the Haissinski equation without any approximations and the Haissinski equation is the exact stationary solution of the VFP equation, we conclude that it is a self-consistent equation for the potential well lengthening applicable to electron storage rings. We emphasize that Eq.~(\ref{eq:PW_Lengthening_from_Haissinski}) is not valid at beam currents above the microwave instability threshold, where the beam energy spread will increase.

Equation~(\ref{eq:Zeff1}) shows that the effective impedance $Z_\parallel^\text{eff}$ is a measure of the average stretching force over the whole bunch. From the impedance viewpoint, Eq.~(\ref{eq:Zeff2}) states that both the real and imaginary parts of the impedance contribute to the bunch lengthening when the density distribution is deformed, although the imaginary part is usually the dominant source.

At $I=0$, the density distribution is Gaussian with $z_c=0$ and $x(0)=1$. The bunch lengthening rate at $I=0$ is given by $x'_0\equiv dx/dI|_{I=0}=\frac{c}{4\pi\sigma_\text{z0}}Z_\parallel^\text{eff}(1)$. Here, the effective impedance $Z_\parallel^\text{eff}(1)$ is given by Eq.~(\ref{eq:Zeff2}) with $z_c=0$ and $\tilde{\lambda}_0(k)=e^{-k^2\sigma_\text{z0}^2/2}$. It only depends on the imaginary part of $Z_\parallel(k)$, suggesting that the inductive part of the impedance solely determines the lengthening rate of the bunch at zero current. The relation between $Z_\parallel^\text{eff}$ and the normalized current $I$ is complicated. Here, we only give the slope at $I=0$ as $dZ_\parallel^\text{eff}/dI|_{I\rightarrow 0}=\frac{2\pi\sigma_\text{z0}}{c}\left[ x_0'^{2} + x'' \right]$ with $x''=d^2x/dI^2$ to be determined.

Further calculations can be carried out when a Gaussian distribution with rms length $\sigma_\text{z}$ and center of mass $z_c$ is used to approximate the Haissinski distribution Eq.~(\ref{eq:Haissinski_equation_explicit}). From Eqs.~(\ref{eq:lossfactor}) and (\ref{eq:Zeff2}), the center of mass and effective impedance are explicitly written as follows:
\begin{equation}
    z_c=\frac{I\sigma_\text{z0}c}{\pi}
    \int_0^\infty dk \text{Re}[Z_\parallel(k)] e^{-k^2\sigma_\text{z}^2},
    \label{eq:zc2}
\end{equation}
\begin{equation}
    Z_\parallel^\text{eff}=-2\sigma_\text{z}^2
    \int_0^\infty dk k \text{Im}[Z_\parallel(k)] e^{-k^2\sigma_\text{z}^2}.
    \label{eq:Zeff4}
\end{equation}
The above quantities can be formulated explicitly for some well-defined impedances in the literature. For more details, the reader is referred to \cite{zhou2023theories}, where it is shown that $Z_\parallel^\text{eff}(x)$ has different scaling laws, depending on the properties of specific impedances. In particular, when the ring impedance is approximated by $Z_\parallel(k)=-ikcL_\text{eff}$, we immediately obtain the cubic equation (\ref{eq:cubic2}) from Eqs.~(\ref{eq:PW_Lengthening_from_Haissinski}) and (\ref{eq:Zeff4}). This justifies Eq.~(\ref{eq:cubic2}) and leads us to conclude that the original Zotter's equation (\ref{eq:cubic1}) should be recast with $\kappa=4\sqrt{\pi}$, as suggested in \cite{smaluk2018impedance}. Meanwhile, the conventional effective impedance and inductance are connected to the newly defined effective impedance $Z_\parallel^\text{eff}$ by
\begin{equation}
    L_\text{eff} = -\frac{1}{\omega_0} \text{Im} \left( \frac{Z_\parallel}{n} \right)_\text{eff}^{m=1}
    = \frac{2\sigma_{z0}}{\sqrt{\pi}c}Z_\parallel^\text{eff},
\end{equation}
and these quantities should be evaluated at zero bunch current.

\section{Bunch lengthening in SuperKEKB}

\begin{figure}
\centering
\vspace{-2mm}
\includegraphics[width=0.85\linewidth]{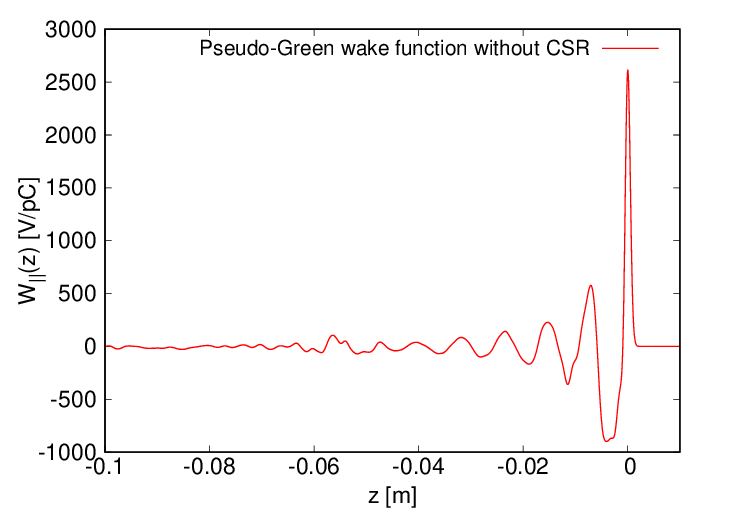}
\includegraphics[width=0.85\linewidth]{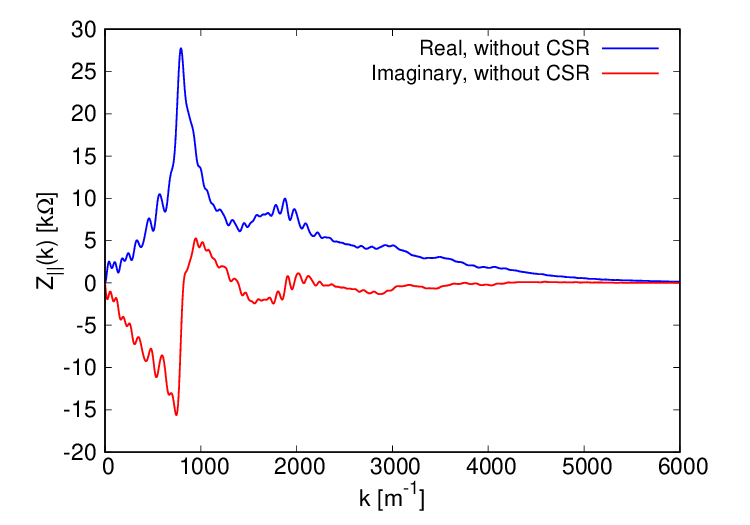}
\vspace{-2mm}
\caption{Pseudo-Green's function wake and corresponding impedance without CSR for SuperKEKB HER.}
\vspace{-2mm}
\label{fig:wakeSKBHER}
\end{figure}

\begin{figure}
\centering
\vspace{-2mm}
\includegraphics[width=0.85\linewidth]{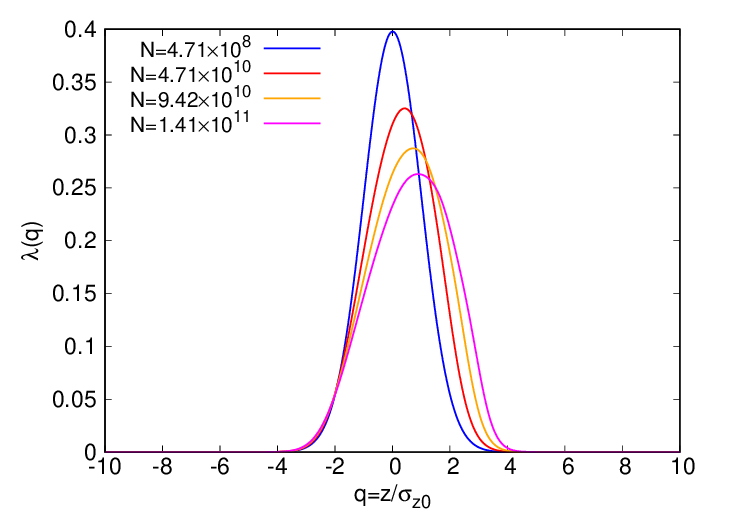}
\vspace{-2mm}
\caption{Haissinski solution for SuperKEKB HER with $N=(0.0471, 4.71, 9.42, 14.1)\times 10^{10}$.}
\vspace{-2mm}
\label{fig:haiDensity}
\end{figure}

\begin{figure}
\centering
\vspace{-2mm}
\includegraphics[width=0.85\linewidth]{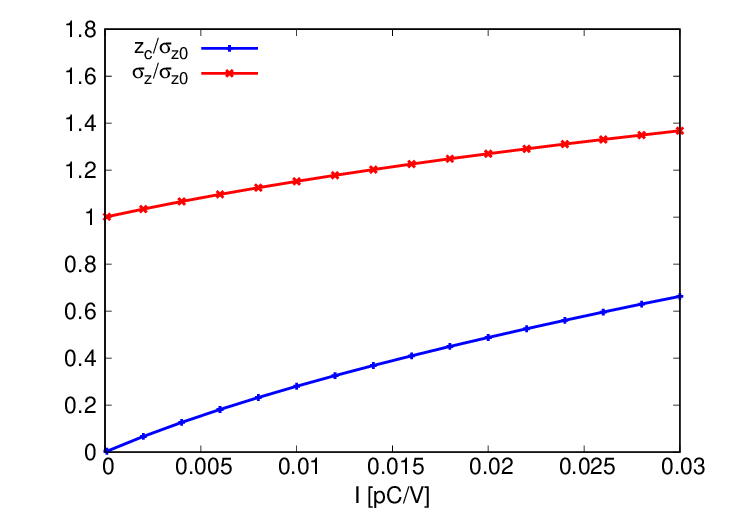}
\vspace{-2mm}
\caption{Bunch centroid and rms bunch length as a function of normalized current for SuperKEKB HER.}
\vspace{-2mm}
\label{fig:sigz}
\end{figure}

\begin{figure}
\centering
\vspace{-2mm}
\includegraphics[width=0.85\linewidth]{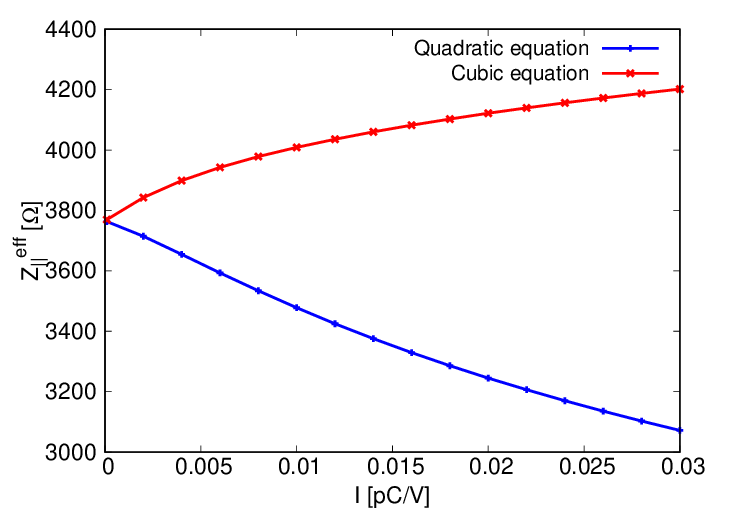}
\vspace{-2mm}
\caption{Effective impedance as a function of normalized current calculated from simulated bunch lengths for SuperKEKB HER.}
\label{fig:ZeffSKB}
\end{figure}

As a specific example, we consider the high-energy ring (HER) of SuperKEKB (the case of LER was discussed in \cite{zhou2023theories}). Longitudinal wakes for various components have been calculated using a Gaussian driving bunch with length $\hat{\sigma}_z$=0.5 mm and summed to create the pseudo-Green function wake~\cite{ishibashi2023impedance} as shown in Fig.~\ref{fig:wakeSKBHER}.
The corresponding impedance shown in the same figure is calculated by Fourier transforming the wake data (indeed, the chirp-Z transform is used to improve the frequency revolution). The decay of impedance data at high frequencies is due to the chosen 0.5mm Gaussian bunch, which acts as a Gaussian window function of $e^{-k^2\hat{\sigma}_{z}^2/2}$. This is justified when the ratio of the nominal length of the bunch $\sigma_\text{z0}$ to the length of the driving bunch $\hat{\sigma}_z$ is very large (for our example, 10). The bunch should not be strongly deformed or micro-bunched, so the high-frequency impedances are not sampled. The beam parameters used to solve the Haissinski equation are beam energy $E$=7.00729 GeV, ring circumference $C$=3016.315 m, slip factor $\eta$=$4.543\times 10^{-4}$, bunch length at zero current $\sigma_\text{z0}$=5.05 mm, momentum spread $\sigma_{\delta0}$=$6.3\times 10^{-4}$, and synchrotron tune $\nu_\text{z0}$=-0.0272. The scaling parameter at $N=10^{11}$ is $I=0.0212$ pC/V. The numerically obtained Haissinski solutions with bunch populations $N=(0.0471, 4.71, 9.42, 14.1)\times 10^{10}$ (corresponding to $I$=(0.0001, 0.01, 0.02, 0.03) pC/V) are shown in Fig.~\ref{fig:haiDensity}.
The lengthening of the bunch and the centroid shift as a function of the normalized current are shown in Fig.~\ref{fig:sigz}.
Using the bunch length data from Fig.~\ref{fig:sigz}, the effective impedance $Z_\parallel^\text{eff}$ 
is calculated by Eq.~(\ref{eq:PW_Lengthening_from_Haissinski}), as shown in Fig.~\ref{fig:ZeffSKB}. In this figure, we also plot the $xZ_\parallel^\text{eff}$ data, which corresponds to Zotter's cubic equation. According to the cubic equation, the inductance impedance dominates the total impedance, and $xZ_\parallel^\text{eff}$ should be a constant or have a weak dependence on the bunch current. Figure~\ref{fig:ZeffSKB} shows that this is not the case for SuperKEKB HER. Therefore, we conclude that Zotter's equation does not apply well to this ring. However, the same analysis for SuperKEKB LER showed that $xZ_\parallel^\text{eff}$ is fairly constant~\cite{zhou2023theories}, suggesting that a pure inductance is good enough to describe the lengthening of the bundle in LER.

\section{Summary}

Zotter’s equation is valid under the following conditions: The longitudinal total impedance of the ring can be well approximated by a pure inductance, and the impact of the real part of the total impedance is negligible. The Haissinski equation, a self-consistent solution of the VFP equation below the microwave instability threshold, has been used to derive a new equation to describe potential-well bunch lengthening. The effective impedance used to describe bunch lengthening has simply been redefined. This new equation is useful for comparing impedance calculations with beam-based measurements in electron storage rings.

\section{ACKNOWLEDGEMENTS}
The author D.Z. thanks many colleagues, including G. Bassi, A. Blednykh, A. Chao, Y. Cai, L. Carver, K. Hirata, R. Lindberg, M. Migliorati, T. Nakamura, K. Ohmi, K. Oide, B. Podobedov, Y. Shobuda, V. Smaluk, M. Tobiyama, and L. Zhang for inspiring discussions on various aspects of impedance issues in electron storage rings.

%
%
\ifboolexpr{bool{jacowbiblatex}}%
	{\printbibliography}%
	{%

} 
%
%


\end{document}